\documentclass[11pt]{article}

\usepackage{amsmath,amssymb,amsfonts}
\usepackage{amsthm}
\usepackage{mathrsfs}
\usepackage{graphicx}
\usepackage{multirow}
\usepackage{xcolor}
\usepackage{textcomp}
\usepackage{hyperref}
\usepackage{lineno}

\usepackage{algorithm2e}
\usepackage{algorithmic}
\RestyleAlgo{ruled}

\usepackage{listings}
\usepackage{subfigure}

\title{Sketching a Space of Brain States}

\author{
Maria Mannone$^{1,2,3,5}$\thanks{maria.mannone@icar.cnr.it; maria.mannone@uni-potsdam.de}
\and
Patrizia Ribino$^{1}$
\and
Peppino Fazio$^{3,4}$
\and
Norbert Marwan$^{2,5,6}$
}

\date{
\small
$^1$ ICAR, National Research Council (CNR), Palermo, Italy\\
$^2$ Institute of Physics and Astronomy, University of Potsdam, Germany\\
$^3$ DSMN, Ca' Foscari University of Venice, Italy\\
$^4$ VSB Technical University of Ostrava, Czechia\\
$^5$ Potsdam Institute for Climate Impact Research (PIK), Germany\\
$^6$ Institute of Geosciences, University of Potsdam, Germany
}

\begin{document}

\maketitle

\begin{abstract}

Brain functional connectivity alterations, that is, pathological changes in the signal exchange between areas of the brain, occur in several neurological diseases, including neurodegenerative and neuropsychiatric ones. They consist in changes in how brain functional networks operate. By conceptualising a brain space as a space whose points are connectome configurations representing brain functional states, changes in brain network functionality can be represented by paths between these points.

Paths from a healthy state to a diseased one, or between diseased states as instances of disease progression, are modelled as the action of the \textit{Krankheit-Operator}, which produces changes from a brain functional state to another. This study proposes a formal representation of the space of brain states and presents its computational definition. References to patients affected by Parkinson's disease, schizophrenia, and Alzheimer–Perusini's disease are included to discuss the proposed approach and possible developments of the research toward a generalisation.

\end{abstract}


\section{Introduction}

\begin{figure}[!h]
    \centering
\includegraphics[width=0.6\linewidth]{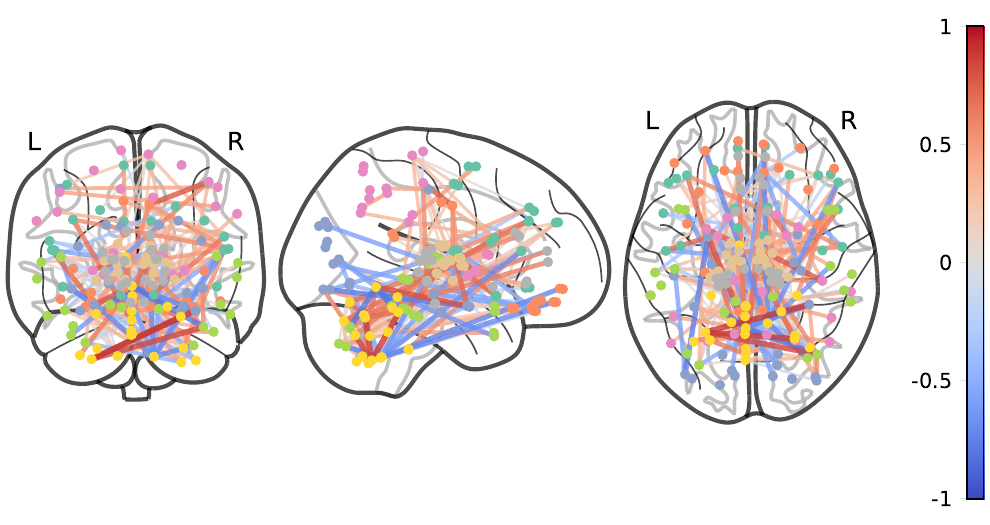}
    \caption{Different perspectives of a human connectome}
    \label{fig:connectome}
\end{figure}

Mind wandering can be a poetic image. What about a multi-dimensional space where brains can actually move through? A space 
whose points are \textit{brain states}. Let us gradually introduce the idea, starting from a construction for each brain, the 
\textit{connectome} \cite{Sporns2005, hagmann_paper}. It is a graph-like representation of the most active brain areas, 
indicated as nodes (often indicated as \textit{hubs} \cite{hubs_brain}), and connected through links (see Fig.\ref{fig:connectome}). 


Structural connectivity is investigated via tractography \cite{SMITH20121924}, whose starting point is the anisotropy of 
diffusion of water molecules, and functional connectivity via functional magnetic resonance \cite{physics_fMRI}, whose 
starting point is the alteration of blood circulation to replenish oxygen to the most active brain areas, at rest, or during 
a given task.

Seeing the brain as an instance of complex networks \cite{structure_complex_network, networks_beyond_pairwise}, the 
alterations of the connections can be related to diseases. This is widely investigated through data analysis in studies 
dealing with the alteration of connectivity \textcolor{black}{in neurological disorders. 
} 
Examples range from disconnection syndrome \cite{disconnection}, to schizophrenia \cite{schizophrenia_heuvel}, to 
neurodegenerative disorders, such as  Alzheimer-Perusini's disease \cite{alzheimer_graph, alzheimer_nature} and Parkinson's disease 
\cite{parkinson_paper, sang2015alteration}, especially including functional connectivity \cite{mosley, morris, shea, 
blandini}.

In an attempt to generalise the approach to neurological disease and pave the way towards integrated systems for healing, as desired
in \cite{mystery2}, a formalism joining matrix algebra, a physics-like approach with operators acting on observables, 
and real-data analysis, the \textit{Krankheit}-Operator (from German \textit{disease}, in short $K$-operator), has been 
proposed \cite{brain_first}. Specific forms of the operator represent the key features of specific neuropsychiatric and 
neurodegenerative diseases.

As a further step in the abstraction, we can imagine a space of connectomes, where each connectome is a point, and 
the transformations between them are paths. 
While several studies investigate how space can be represented within the human brain \cite{buz}, we are interested in 
finding how a brain can be represented within a conceptual space, as a parameter space. The closer attempt to such an endeavour 
was proposed in \cite{fard}. The authors, drawing upon Waddington’s epigenetic landscape for cell development, propose a 
Hopfield network formalism to build up an attractor model of disease progression based on networks of proteins or genetic 
correlation. 
Disease progression is modelled through the reference to curves within a space, where each point is a specific network 
configuration; however, the focus is more on genetics \cite{fard}. Normal state and diseased states are space regions 
considered as attractors.
In their 3D representation, the axes contain the first principal component, the second principal component, and the energy. The presented 
examples include Parkinson's disease, glioma, and colon cancer. 

Our representation, developed independently from this study, also contains states within a space and subspaces for healthy 
and diseased states, but it is built in a completely different way, considering the brain connectome and encoding information 
on brain areas into the three axes.
As another difference, we shift the attention from attractors within the space to the operator leading time evolution, in 
particular, considering disease time evolution, that is, the $K$-operator \cite{brain_first}.
Other studies focus on brain diseases, seeing them as attractors \cite{attractors}.




In this article, \textcolor{black}{inspired by the phase space theory \cite{reichl} where the set of all possible physical 
states of a system corresponds uniquely to points in the phase space}, we propose a Brain Space where points represent 
specific brain states and whose paths are the transitions from a brain state to another one, as mappings between them. 
Choosing to describe a brain as its connectome, the points of the space are specific connectomic configurations, and the 
paths are transformations of connectomes \textcolor{black}{, i.e., changes in brain-network functionality.}


Before moving forwards, we present in Table \ref{acronyms} a little glossary of the most used acronyms in this paper.
\begin{table}[ht!]
\centering
\begin{tabular}{ll}
\hline
Acronym & Complete Name \\
\hline
fMRI & Functional magnetic resonance imaging \\
DICOM & Digital Imaging and Commeunications in Medicine \\
NIfTI & Neuroimaging Informatics Technology Initiative \\
ROI & Region of Interest (of the brain) \\
ADNI & Alzheimer's Disease
Neuroimaging Initiative \\
PPMI & Parkinson’s Progression Markers Initiative \\
COBRE & Center for Biomedical Research Excellence \\
FU & Follow-Up \\
AD & Alzheimer(-Perusini)'s disease \\
PD & Parkinson's disease \\
thal\_MGN\_R & Medial
Geniculate Right\\
thal\_LGN\_R & Lateral Geniculate Right \\
\hline
\end{tabular}
\caption{Table with the most used acronyms in this article.}\label{acronyms}
\end{table}

\textcolor{black}{Our proposed approach is theoretical, and the selected case studies are used here to illustrate the method. They are not exhaustive, but they help instantiate the definitions, define the methodology, the advantages, and the limitations.}

The rest of the paper is organised as follows. We sketch our theoretical idea in Section \ref{theory}, and we propose a 
possible implementation of it in Figure \ref{data-driven_visualization}, discussing the results in Section \ref{results}. 
Possible research developments are listed in the Conclusions. 

\section{Theoretical view}\label{theory}

\begin{figure}[t]
\centering
\includegraphics[width=\linewidth]{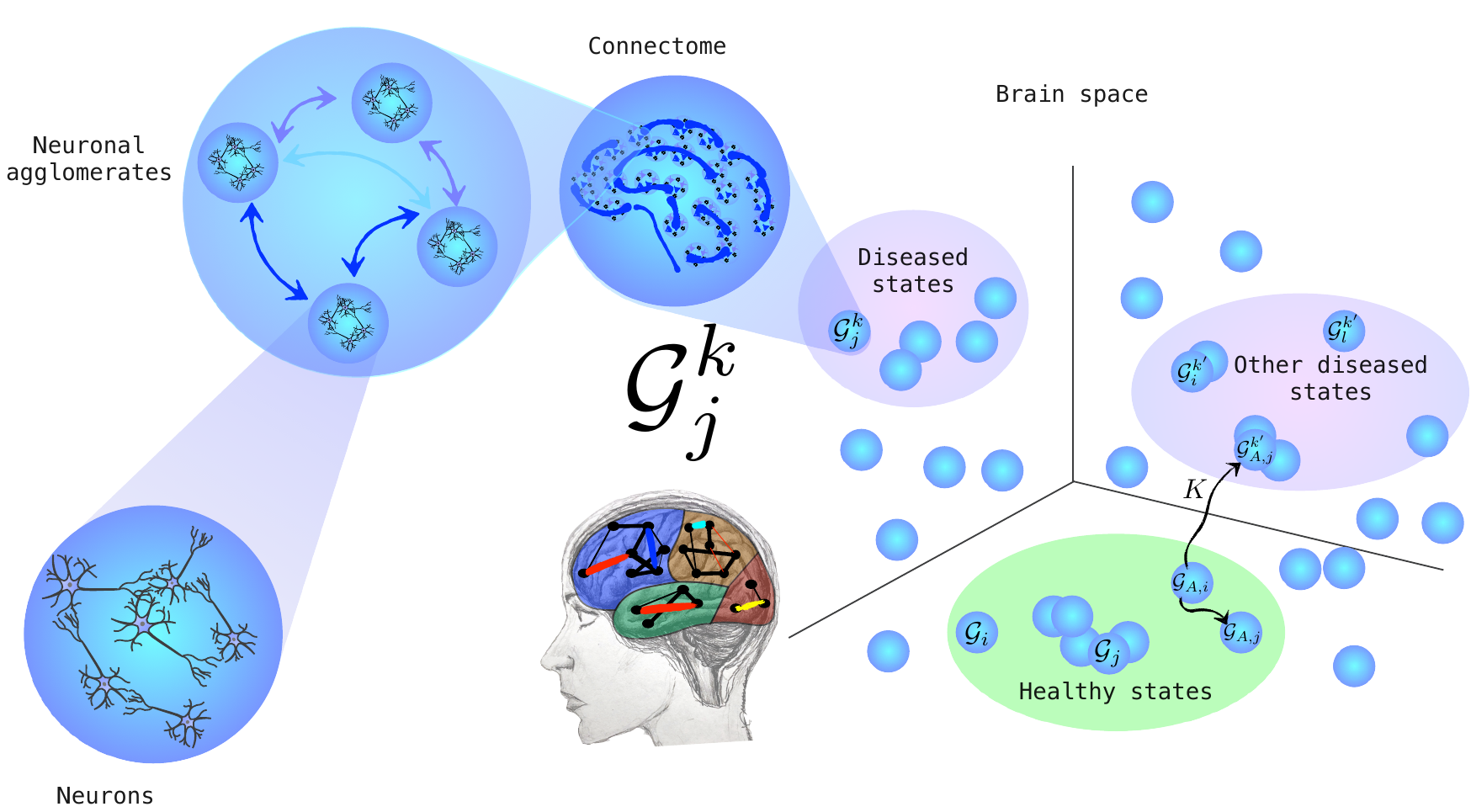}
\caption{Left: pictorial representation of the brain as a nested network, from neurons, to neuronal agglomerates, to the 
connectome. Right: Space of brain states, where each connectome is a point, and the transformation of a specific brain over 
time can be represented as a path from one state to another. \textcolor{black}{Bottom right: the path from a brain of person A from state $\mathcal{G}_{A,i}$ to state $\mathcal{G}_{A,j}$ lies within the healthy states, while the pathological evolution brings the brain towards the state $\mathcal{G}^{k'}_{A,j}$ within one of the diseased states' subspaces. (Drawing and graphics by Maria Mannone).}}
\label{fig4}
\end{figure}

We propose a \textbf{Brain Space as a phase space}, where each point is a connectomic state. But we know that each brain is 
in itself a nested network, from neurons, to neuronal agglomerates, to the connectome (Figure \ref{fig4}). A transformation 
of the connectome over time is a path between points in the Brain Space.
The size of the connectome depends upon the chosen number of nodes, that is, the number of regions of interest (ROIs) the 
brain can be divided into. \textcolor{black}{A brain atlas \cite{AAL3_ref1, MSDL, oxford} defines the shape and location of 
brain regions in a common coordinate space. Hence,} the degree of resolution of brain analysis influences the choice of brain 
atlas: the more the regions of interest, the higher the number of points of the connectome.

Let $N$ be the number of ROIs in a chosen atlas and let us consider retaining all information, then the number of degrees of 
freedom in our brain space will be $N$. \textcolor{black}{From phase space theory}
if we consider \textcolor{black}{all possible 
values of the} \textit{position} \textcolor{black}{(i.e., the state in which a brain is)}
and \textit{momentum} (i.e., the movement from one state to another one), then the number of degrees of freedom doubles to 
$2N$, resulting in a $2N$-dimension brain space.

What are \textit{position} and \textit{momentum} for the connectome? 
\textcolor{black}{Since a connectome is a dynamical model,} the position indicates where a connectome is at time $t$, and the 
momentum determines how it is changing from one state to another. 


The transition from one state to another is a normal process due to learning (with the consequent development of brain 
areas), feeling emotions, or ageing. A normal time evolution is described with paths within the ``healthy states'' subspace, \textcolor{black}{see the bottom right side of Figure \ref{fig4}, where the ``healthy evolution'' brings a brain of a person generically indicated as A, in a healthy state $\mathcal{G}_{A,i}$, towards another healthy state $\mathcal{G}_{A,j}$.} 


\textcolor{black}{On the other hand, } 
abnormal dynamical processes concern the transitions under the effect of 
\textcolor{black}{functional neurologic
disorder} onset, or exceptionally overwhelming emotions or trauma. A brain point can move towards states outside the healthy-state 
subspace, \textcolor{black}{reaching a diseased state; see the path for the brain of a generic patient A toward $\mathcal{G}^{k'}_{A,j}$, under the effect of the action of $K$, on the right side of Figure \ref{fig4}.} 

\begin{figure}[h!]
\centering
\includegraphics[width=0.6\linewidth]{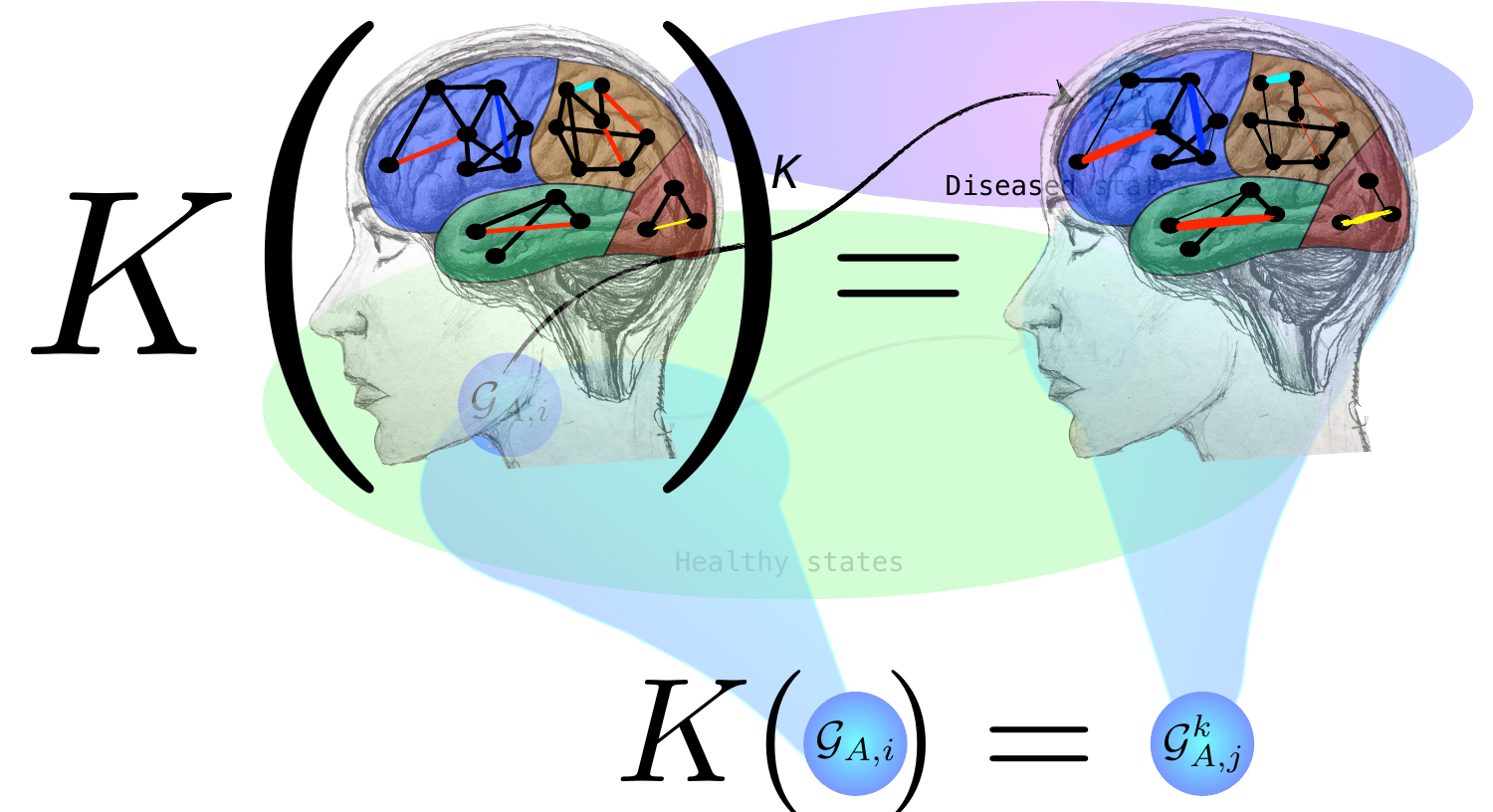}
\caption{Pictorial representation of the $K$-operator on a brain. Drawing by Maria Mannone.}
\label{fig3}
\end{figure}

\textcolor{black}{In fact,} such a process develops under the action of $K$ (see Fig.~\ref{fig3}), modelled as an operator altering the weights of links 
in the connectome, reproducing the onset or progression of specific neurological disorders \cite{brain_first, PD_paper}, as in 
Eq.~\eqref{action_K}:


\begin{equation}\label{action_K}
K(t)\mathcal{G}^k(t)=\mathcal{G}^k(t+1).
\end{equation}

In Eq. \eqref{action_K}, $K$ may act on a healthy connectome leading to a diseased one or on an already-diseased brain, 
letting the disease worsen.
Hence, the action of $K$ leads a brain along a path within subspaces of specific diseases or brings a brain out of the 
healthy subspace or even back and forth, as in epilepsy or mood disorders.
Thus, the paths in the brain space are governed by ``simple dynamics'' if the states and the paths between them remain within 
the healthy subspace.
\textcolor{black}{That is, the brain dynamics can keep the states of a brain gravitate within a basin of attraction. The precise shape of the dynamics depends upon the changes of the brain: normal ageing, continuous learning, new skills development all contribute to the shaping of the brain and functional connectivity. While also the anatomic connectivity is affected by pathological and, up to a certain extent, also by normal processes (as learning how to play a musical instrument), in this study we focus on the functional one. An ideal ``normal state'' considering age range can be treated as a probabilistic attractor, towards which the brain states gravitate to.}
\textcolor{black}{If more disorders are present in the same patient, the matrix elements of the $K$-operator for these disorders are also involved. A precise discrimination can be performed when we compare  the $K$ obtained from patients with comorbidities with this $K$s obtained from patients affected by only one of these diseases.}

The paths are governed by the $K$-operator if the states are outside or moving from the healthy 
subset to the outside. 
A process bringing a brain back to the healthy subspace 
is a healing process. The ideal healing is the inverse of the $K$-operator:  $K^{-1}$. A sequence of healing steps for a 
progressive healing process is described in \cite{frontiers_paper}, as well as the definition of a \textit{missing therapy} 
from the ideal healing and the already-existing, partial healing strategies. 

\textcolor{black}{The $K$-operator is a nonlinear operator, but some linear approximations have been proposed. It is time-dependent; in \cite{brain_first}, a precise functional, t-dependent shape of the operator was proposed for its elements after the characteristics of some specific disorders. In other studies \cite{PD_paper}, only specific numerical forms of the $K$-operators have been computed, between two time points.}

\textcolor{black}{The key idea is the theoretical definition of a space of brain states. The use of multi-dimensional scaling (MDS), that will be discussed later, is only a possible way to compress the information concerning the ROIs. Thus, MDS is only instrumental. The key point is the definition of a conceptual framework to represent the transformation of brain states over time, both within the healthy domain, and between healthy and pathological states, which is characteristic of diseases such as epilepsy or bipolar disorders. The $K$-operator is a novel idea that starts being developed. Our wish is that it can also be further adopted and explored by different researchers. However, in this study, the concept of $K$ is merely used as label for the trajectory of a brain across states including diseased states, that is, to characterise its pathological time evolution.}

Summarizing, we can state that 
\textbf{a brain is \textit{healthy} if all its states throughout time remain within the healthy states subspace in a stable 
way. A diseased brain heals if its states move from the disease's subspaces towards the healthy states' subspace and then 
remain there stably.}
\textcolor{black}{Remaining within the healthy-state subspace does not mean that the brain is not changing: it means that it changes in a normal, non-pathological way. It is the case of the growing process, from childhood to adulthood, and progressively with healthy ageing. Also learning shapes the brain, strengthening some connections and enlarging some brain areas. A classic example is constituted by the brain changes in musicians \cite{brain_music, musicophilia}. Thus, they impact both the anatomic connectivity, as well as the functional connectivity.}


\section{A two-stage approach to compute a Brain Space}
\label{data-driven_visualization}
\textcolor{black}{
Here, we describe the two-stage approach to compute the Brain Space. The first stage allows for obtaining a single brain state 
points of the space starting from the fMRI images, while the second one allows for the construction of the whole brain 
space.}

\textcolor{black}{As said before, the points of the proposed brain space are connectomes. }
By focusing on fMRI-derived data from some real patients, we obtain connectomes as a connectivity matrix for each brain 
according to the steps described in the following pseudocode.

\begin{algorithm}[ht!]
\caption{Brain States  Computation}
\KwData{fMRIs as DICOM files}
\KwResult{Brain Space points}
\BlankLine
\textbf{Step 1: fMRI conversion}\\
\Indp
Convert the DICOM files to NIfTI format\;
\Indm
\BlankLine
\textbf{Step 2: Atlas selection}\\
\Indp
Choose a medical atlas to be used as a mask to group matrix elements of NIfTI
within specific brain areas\;
\Indm
\BlankLine
\textbf{Step 3: Time series extraction}\\
\Indp
Extract time series for each ROI\;
\Indm
\BlankLine
\textbf{Step 4: Connectivity matrix computation}\\
\Indp
Compute the connectivity matrix as a correlation matrix\footnote{\textcolor{black}{In this paper, we used the Pearson 
correlation. However, other correlation methods can be used.}}, whose elements are the correlation values between pairs of 
ROIs averaged over time\;
\Indm
\BlankLine
\textbf{Step 5: Matrix labelling}\\
\Indp
Label each matrix as $\mathcal{G}^k_j(t)$ for disease $k$ and patient $j$ at the time visit $t$\;
\Indm
\BlankLine
\end{algorithm}


For instance, the brain matrix of a patient at the first visit and the brain matrix of the same patient at the first 
follow-up 
will be two different points in the space of brains, and the transition from baseline to follow-up will be a path in the 
brain space between these two points. However, no absolute time is considered: e.g., the brain state of another patient, 
whose baseline measure was obtained six months before the baseline of the first patient, will just be another point in the 
space. In this way, we obtain the single points of our brain space, and that will be used to compute the values on the axes 
via a \textit{multi-dimensional scaling} (MDS) as described in the following. \textcolor{black}{The technique to compute MDS is standard, however here the novelty is represented by the overall framework of brain spaces, and by the meaning of dimensions in this study. The MDS dimensions are orthogonal, and they are computed as the weighted sum of the 160 regions of interest (ROIs) of the human brain for which all the considered brain matrices are non-empty. The weights are computed according to the impact of the Frobenius distance between corresponding elements of the matrices representing functional brain connectivity in each considered brain state. The higher the weight, the more important a pair of ROIs is for distinguishing between different brain states. The correlation of a single ROI to the axis takes into account its relative impact on the different pairs, as it will be further explained later.}

Let us now describe how a brain space can be obtained, how we can shape it as a 3D shape for visualisation, and what meaning 
can be assigned to the axes. 
Through the MDS algorithm, arbitrarily selecting $3$ as the number of dimensions, we can automatically arrange the matrix 
points in the space. The closer the points, the more similar the matrices (and thus the corresponding brain states). 
To this aim, here we consider element-wise similarity (with Euclidean distance) of the matrices. \textcolor{black}{In particular, concerning the distance metric, we finally decided to focus on the Frobenius distance.}
The main steps to represent the brain space are illustrated in the following pseudocode.










\begin{algorithm}[ht!]
\caption{Brain Space Representation}
\KwData{$\mathcal{G}^k_j(t)$ for disease $k$ and patient $j$ at the time visit $t$ }
\KwResult{3D visualization of Brain Space}
\BlankLine
\textbf{Step 1: ROI selection}\\
\Indm
\BlankLine
\textbf{Step 2: Space points distribution}\\
\Indp
Distribute space points according to element-wise similarity\;
\Indm
\BlankLine
\textbf{Step 3: MDS scaling}\\
\Indp
Perform multi-dimensional scaling according to the space dimension\;
\Indm
\BlankLine
\textbf{Step 4: Axes value computation}\\
\Indp
Compute the values of the axes as contributions of the ROI with weights automatically chosen according to the inputted brains\;
\Indm
\BlankLine

\textbf{Step 5: Plotting}\\
\Indp
Plot the Brain Space\;
\Indm
\BlankLine

\textbf{Step 6: ROI contribution}\\
\Indp
Print list of the ROIs and their correlation to each axis\;
\Indm
\BlankLine

\textbf{Step 7: Better ROIs selection}\\
\Indp
Finding which ROIs help better differentiate between brain states\;
\Indm
\BlankLine
\end{algorithm}




\textcolor{black}{No ROIs were removed as uninformative; the only regions actually removed are those missing from one or more matrices of the selected brains. But this step was preliminary to the overall analysis, and thus 160 ROIs out of 170 ROIs in the Anatomic Automatic Labeling 3 (AAL3) atlas were considered.}
Finally, we provide the details for performing the MDS in step 4. For each axis, MDS performs a weighted sum of the ROIs, where the most influential ones have higher weight (that is, higher correlation as an absolute value with the axis).

\begin{algorithm}[ht!]
\caption{Projection of 3D Matrices according to their similarity}
\KwData{Matrix of distances $D$, labels $L$, names of the matrices $N$, dictionary of visualised indices $S$}
\KwResult{Interactive 3D visualization of matrices}
\BlankLine
\textbf{Step 1: MDS projection}\\
\Indp
Run MDS on $D$ to get a 3D representation $\mathbf{P}$\;
\Indm
\BlankLine
\textbf{Step 2: initialization of the 3D graph}\\
\Indp
Create a new empty 3D figure $fig$\;
\Indm
\BlankLine
\textbf{Step 3: addition of matrices to the graph}\\
\Indp
Add to $fig$ a scatter plot 3D, by using $\mathbf{P}$, with labels $N$ and colours based on $L$\;
\Indm
\BlankLine
\textbf{Step 4: addition of complex hulls for selected groups}\\
\Indp
\If{$S \neq \emptyset$}{
    assign a list of distinct colours $C$\;
    \ForEach{group $g$ in $S$}{
        select points $\mathbf{P}_g$ corresponding to $g$\;
        compute the complex hull $H_g$ of $\mathbf{P}_g$\;
        extract vertices $V_g$ of $H_g$\;
        add to $fig$ a mesh 3D for $H_g$ with colour from $C$\;
        add to $fig$ a scatter plot 3D for the points in $g$ with colour from $C$.
    }
}
\end{algorithm}

\section{Experimental Results}\label{results}
In this section, we provide a computation of a Brain Space obtained from a set of patients. In particular, to 
highlight differences between brain functional alterations due to different diseases, we choose patients from 
different healthcare datasets relating to different neurological disorders as listed in Table 
\ref{list_patients}. \textcolor{black}{We chose states of patients with healthy brains or with diseased brain, with features representative of the respective disorders. In particular, we focus on patients individually investigated for preliminary applications of the $K$-operator for specific disorders.} Moreover, we chose the Anatomic Automatic Labeling 3 (AAL3) atlas \cite{AAL3_ref3} and 
selected the 160 ROIs within all the fMRIs of the considered patients.

\begin{table}[ht!]
\centering
\resizebox{\textwidth}{!}{
\begin{tabular}{lcrcccc}
\hline
label & disease & ID patient & time & age & sex & dataset \\
\hline
AD & AD & 002\_S\_5018 \cite{limbic} & baseline & 73 & male & ADNI 2 \\
PD & PD & 100878 \cite{PD_paper} & baseline & 67 & male & PPMI \\ 
schizo & schizophrenia  & sub-A00015518 \cite{PD_paper} & baseline & 60 & male & COBRE \\
normal & -- & 101195 \cite{PD_paper} & baseline & 74 & male & PPMI \\
AD\_fem & AD & 019\_S\_5019 \cite{limbic} & baseline & 63 & female & ADNI 2 \\
AD\_fem\_FU & AD & 019\_S\_5019 \cite{limbic} & follow-up & 63 & female & ADNI 2 \\
AD\_patC & AD & 006\_S\_4153 \cite{limbic} & baseline & 79 & male & ADNI 2 \\
AD\_patC\_FU & AD & 006\_S\_4153 \cite{limbic} & follow-up & 79 & male & ADNI 2 \\
PD\_patB & PD & 100006 \cite{PD_paper} & baseline & 56 & female & PPMI \\
PD\_patB\_FU & PD & 100006 \cite{PD_paper} & follow-up & 56 & female & PPMI \\
normal\_F & -- & 018\_S\_4399 \cite{limbic} & baseline & 78 & female & ADNI 2 \\
PD patient$^{\ast}$ & PD & 101050 & baseline & 50 & female & PPMI \\
\hline
\end{tabular}
}%
\caption{List of patients in Figure \ref{fig5_intro} and \ref{fig6_one_more_patient2}. 
($^{\ast}$ is not present in Figure \ref{fig5_intro}, but was included in Figure \ref{fig6_one_more_patient2}). 
AD stands for Alzheimer-Perusini's disease, PD for Parkinson's disease. \textcolor{black}{The references indicate preliminary studies on the $K$-operator where these patients have been considered.}}
\label{list_patients}
\end{table}



\begin{figure}[h!]
\centering
\includegraphics[width=\linewidth]{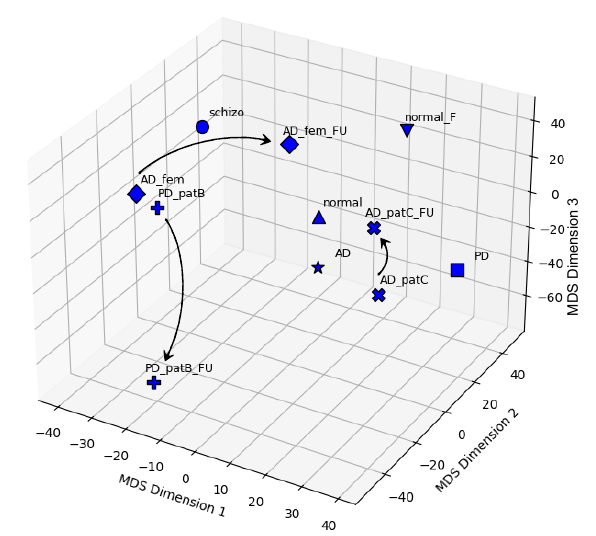}
\caption{A quantitative example of brain space for a selection of patients. For the explanation of the 
points' labels, see Table \ref{list_patients} (right). The labels contain a weighted sum of the regions of 
interest, automatically computed (Table \ref{labels_ROIs_MDS}). \textcolor{black}{The states belonging to the 
same patients are identified with the same symbol. The arrows, indicating time evolution, are added as 
post-processing.}}
\label{fig5_intro}
\end{figure}

In Figure \ref{fig5_intro}, 
we can see the Brain Space generated for the patients in Table \ref{list_patients}.

\begin{figure}[ht!]
\centering
\includegraphics[width=\linewidth]
{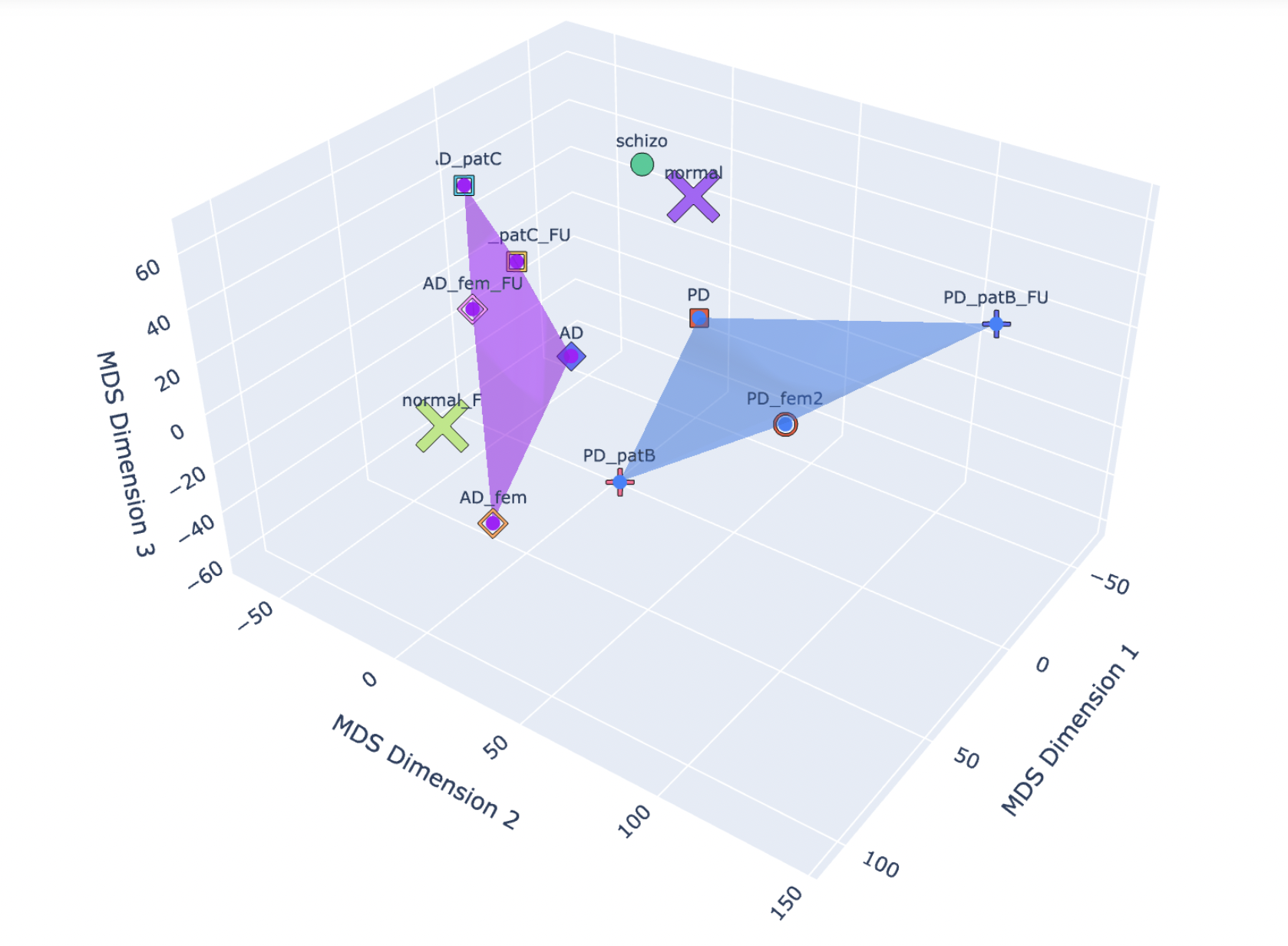}
\caption{A quantitative example of brain space for the selected patients, where \textcolor{black}{we also use 
 the same symbol for the same patient, with the exception of the two normal ones, with the same symbol, 
an $\times$ (limitation of the available marker shapes in 3D in Python). Here: Euclidean dissimilarity.} 
}
\label{fig6_one_more_patient2}
\end{figure}

On the other hand,
Fig.~\ref{fig6_one_more_patient2} shows the simplex (violet) of Alzheimer-Perusini's patients and the simplex 
(blue) of Parkinson's patients. As we can see, there is a clear separation between these disease subspaces.


As concerns the schizophrenic patient, he is distant from the PD patients and even more distant from the 
follow-up of one of the PD patients. 
This makes sense since the evolution of PD involves a defect of dopamine neurotransmitters due to damage in 
the substantia nigra, and a consequent deficit in dopamine is ``opposed'' to the excess of dopamine occurring 
in the same region for schizophrenic patients \cite{VANHOOIJDONK202357}.\footnote{Pathways 
of dopamine are disrupted in different ways in schizophrenic brains. According to some authors, the excessive 
dopamine activity in the mesolimbic neurons is caused by an abnormally low prefrontal dopamine activity, 
leading to cognitive symptoms \cite{davis}.}



The axes of Figures \ref{fig5_intro} and \ref{fig6_one_more_patient2} contain the synthetic description of 
MDS dimensions 1, 2, and 3. 
The MDS algorithm automatically computes a weighted sum of the contribution of each region of interest for 
each axis in terms of correlation values.
The higher the ROI absolute value, the higher the impact of the ROI in the space separation between 
points.

\textcolor{black}{It is worth noting that 
MDS is a dimension-reduction technique designed to project high-dimensional data down to lower dimensions 
while preserving relative distances between observations. Since in our example we reduce $160$ ROIs to 
obtain a 3D representation of the Brain Space, the more points we add to the Brain Space, the more the axes values are refined.
}

Table \ref{labels_ROIs_MDS} shows the first five more impacting ROIs on each axis, before (left) and after 
(right) the addition of the last patient of Table \ref{list_patients}. With a larger population, the 
classification could become more precise. 
Nevertheless, with this simple example, we also draw information on the feasibility of our brain state 
definition from a purely theoretical approach to a computational one.

\begin{table}[ht!]
\centering
\resizebox{\textwidth}{!}{
\begin{tabular}{ccc|ccc}
\hline
\multicolumn{6}{c}{Dimension 1 (axis $x$)} \\
\hline
ROI & correlation & ROI name & ROI & correlation & ROI name \\
\hline
71	& 0.739	& Precuneus\_L & 60 & 0.702 & fusiform\_R \\
30	& -0.724 & OFCpost\_R & 72 & 0.696 & precuneus\_R \\
16	& 0.724	& Supp\_Motor\_Area\_R & 66 & 0.688 & parietal\_inf\_R \\
66 & 0.688 & parietal\_inf\_R & 2 & 0.685 & precentral\_R \\
64	& 0.719	& Parietal\_Sup\_R 
& 140 & 0.680 & thal\_LGN\_R \\	
\hline
\multicolumn{6}{c}{Dimension 2 (axis $y$)} \\
\hline
ROI & correlation & ROI name & ROI & correlation & ROI name \\
\hline
102	& 0.866	& Cerebellum\_4\_5\_R & 20 & 0.868 & frontal\_superior\_med\_R \\
84	& 0.833	& Heschl\_R & 14 & 0.838 & rolandic\_oper\_R \\
56 & 0.794 & Occipital\_Mid\_R & 66 & 0.835 & parietal\_inf\_R \\
86 & 0.750 & Temporal\_Sup\_R & 10 & 0.830 & frontal\_inf\_tri\_R \\
10	& 0.738 & Frontal\_Inf\_Tri\_R & 6 & 0.824 & frontal\_mid\_2\_R \\
\hline		
\multicolumn{6}{c}{Dimension 3 (axis $z$)} \\
\hline
ROI & correlation & ROI name & ROI & correlation & ROI name \\
\hline
106	& -0.877 & Cerebellum\_7b\_R & 142 & 0.824 & thal\_MGN\_R \\
91	& -0.875 & Temporal\_Pole\_Mid\_L & 89 & 	-0.804 & temporal\_mid\_L \\
89	& -0.837 & Temporal\_Mid\_L & 91 & -0.800 & temporal\_pole\_mid\_L \\
85	& -0.820 & Temporal\_Sup\_L & 106 & -0.798 & cerebellum\_7\_R \\
99	& -0.809 & Cerebellum\_3\_L & 55 & -0.794 & occipital\_mid\_L \\
\hline
\end{tabular}
}%
\caption{The first five ROIs more heavily affecting (in absolute value) the axis content of the brain space, before (left) and after (right) the inclusion of the last patient in Table \ref{list_patients}.}
\label{labels_ROIs_MDS}
\end{table}

\textcolor{black}{The MDS are computed after the pairs of ROIs, the most influential to distinguish between brain states according to the $\mathcal{G}$ matrices. They can also be related to the single ROIs most influential in those pair variations. 
Our reported ROI correlations are still exploratory, but the analysis can be enriched with the features directly derived from the connectivity matrices.
In fact, the individual impact of ROIs on an MDS axis is obtained via the sum of the contributions of all connections involving that ROI. And it includes a 1/2 factor, to avoid counting a pair of ROIs twice. By considering the impact of the ``original'' pairs of ROIs, we can listen the first of them according to their impact, as shown in Table \ref{list_pairs}, directly for the case study with all patients.}
\begin{table}[ht!]
\begin{tabular}{ccc}
\hline
1st ROI & 2nd ROI & Variance \\
\hline
115 & 116 & 0.33 \\
\textbf{74} & \textbf{78} & 0.33 \\
114 & 116 & 0.32 \\
46 & \textbf{70} & 0.31 \\
104 & 115 & 0.30 \\
49 & \textbf{70} & 0.30 \\
113 & 116 & 0.30 \\
106 & 115 & 0.30 \\
\textbf{98} & 116 & 0.29 \\
57 & \textbf{86} & 0.29 \\
45 & \textbf{70} & 0.29 \\
\textbf{61} & \textbf{99} & 0.29 \\
100 & 116 & 0.29 \\
105 & 116 & 0.29 \\
151 & 152 & 0.28 \\
104 & 116 & 0.28 \\
\textbf{99} & 115 & 0.28 \\
61 & 106 & 0.28 \\
\textbf{57} & 115 & 0.28 \\
46 & \textbf{61} & 0.28 \\
\hline
\end{tabular}
\caption{The most influential pairs of ROIs. The bold indicates the ROIs also present in the list of the first five most influential individual ROIs in each axes, or very close ones.}\label{list_pairs}
\end{table}

			

\clearpage
\newpage

\textcolor{black}{The MDS distribution is sensitive to the characteristics of inputted brain states. In fact, a progressive refinement can be achieved with the addition of more patients.}

To complete the discussion of the brain space, we also give a quantitative representation of a path. Within 
the brain space, the progression of the disease is an arrow, labelled as the $K$-operator, transforming, for 
instance, the brain state of a Parkinsonian patient at the baseline (for example, PD\_patB) to the state at 
the first follow-up (for example PD\_patB\_FU).
If we consider a space of a number of dimensions equal to the number of ROIs, thus $160$ in the present case, 
then the brain-state path from baseline to follow-up is perfectly described by the $K$-operator computed 
between the two states and for that choice of atlas. However, in these compressed dimensions via a 
multi-dimensional 
scaling, some ROIs will contribute more than others, depending on the other states in space, see 
Table \ref{list_patients}. Then, the path under consideration will be described by a compressed version of 
the $K$-operator.
By extracting the coordinates of the points PD\_patB and PD\_patB\_FU, we can write:
\begin{equation}
\begin{split}
K|_\text{space\,MDS,\,3D}:\,&PD\_patB \rightarrow PD\_patB\_FU \\&\Rightarrow[-34.25, -15.81, 
-3.611]\rightarrow[-13.48,-47.21,-69.99].
\end{split}\end{equation}
Hence, the \textbf{path from PD\_patB to PD\_patB\_FU can be computed as the percentage of variation of each 
of these ROIs}.
For the MDS Dimension 1 we have 60.64\%, for Dimension 2 we have 198.61\%, and finally for Dimension 3 we get 
1838.24\%. These percentage variations have to be reported on the five most influential ROIs and their 
correlation with the axes from Table \ref{labels_ROIs_MDS}. Limiting ourselves to the first ROIs for each 
axis, we can assess that the strongest alteration occurs for the third MDS dimension, and thus mostly impacting 
the first ROI of the list, cerebellum 7b right or the thalamus geniculate (considering the left or right side 
of Table \ref{labels_ROIs_MDS}), followed at large by the second MDS dimension with cerebellum 4R or frontal 
superior medial R, and the first MDS dimension, whose greater weight is constituted by precuneus left or fusiform right. However, due to the two orders of magnitude of the change in the third MDS dimension, also the other four ROIs that are impacting that axis are worthy of notice (temporal mid-left, temporal pole mid-left, cerebellum 7 right, and occipital mid-L).

\begin{figure}[ht!]
\centering
\includegraphics[width=1.1\linewidth]{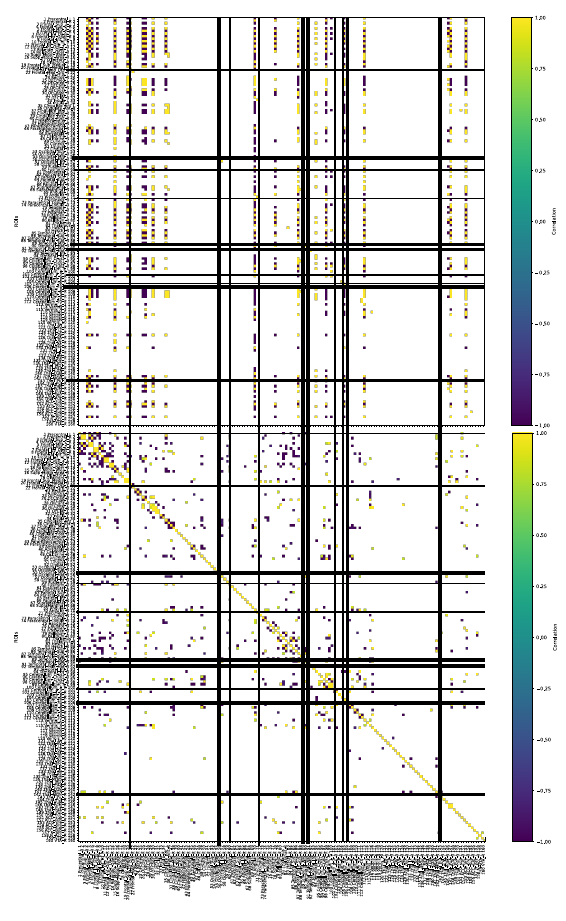}
\caption{ $K$-operator for the disease progression of patient B (Table \ref{list_patients}), from the 
baseline to the first follow-up. The thicker lines indicate the first ROIs more correlated with the third MDS 
dimension, where the change from baseline to follow-up is higher. (Original figure taken from \cite{PD_paper})}
\label{patient_B_highlighted}
\end{figure}

By looking at Fig.~\ref{patient_B_highlighted} taken from \cite{PD_paper} representing the $K$-operator for the disease progression of patient B and highlighting the mentioned regions (with thicker lines for the first ROIs of the 
third dimension), we check that they correspond to specific clusters of (mostly blue, i.e., negative) 
points in the $K$-operator for the $160$ ROIs.

\textcolor{black}{We can relate the coordinates of the two brain states connected by the considered $K$-operator to the most influential pairs of ROIs. In particular, we can focus on the third axis, the one where the higher change is found. We thus obtain the list of Table \ref{K_pairs}. In particular, we notice the pairs:
\begin{itemize}
\item 15-87 supplementary motor area, and temporal pole: superior temporal gyrus;
\item 5-6, both middle frontal gyrus;
\item 27-87 anterior orbital gyrus, and temporal pole: superior medial frontal gyrus,
\end{itemize}
and in particular the ROIs:
\begin{itemize}
\item 3: superior frontal gyrus,
\item 35: anterior cingulate,
\item 1: precentral gyrus,
\item 37: middle cingulate gyrus,
\item 84: Heschl's gyrus,
\item 70: angular.
\end{itemize}
From the evaluation of the complete $K$-operator (Figure \ref{patient_B_highlighted}), we can notice the impact of the nearby pairs of ROIs 89-21, 90-21, near to 15-87 and 27-87, and we can also notice, from visual inspection, the importance of the agglomerates around 5-6, and the small agglomerates in correspondence of 71-35. The highlighted areas, near to the regions of Table \ref{K_pairs}, are the following: 
\begin{itemize}
\item 89: middle temporal gyrus,
\item 21: superior frontal gyrus,
\item 27: anterior orbital gyrus,
\item 71: precuneus, and
\item 35: anterior cingulate.
\end{itemize}
}

\begin{table}[ht!]
\centering
\begin{tabular}{ccc}
\hline
ROI-ROI pair & connectivity change & axis estimated contribution \\
\hline
15 - 87 &  1.78 & -1.549 \\
5 - 6 & -1.76 & 1.535 \\
27 - 87 & -1.74 & 1.517 \\
3 - 6 & -1.74 & 1.511 \\
6 - 35 & -1.71 & 1.489 \\
1 - 87 & 1.71 & -1.485 \\
6 - 37 & -1.71 & 1.484 \\
84 - 87 & -1.70 & 1.483 \\
35 - 87 & 1.70 & -1.481 \\
37 - 70 & -1.70 & 1.478 \\
\hline
\end{tabular}
\caption{The most influential pairs of ROIs concerning the third axis, MDS dimension 3, where the higher change is highlighted for the considered example (patB, from baseline to the first follow-up). The values in the last column decrease (in absolute value) from the first to the last column. This is more evident if more pairs are printed.}
\label{K_pairs}
\end{table}

\textcolor{black}{Summarising, the ``classic'' strategy for the computation of the $K$-operator is: ROIs$\rightarrow\mathcal{G}$-matrices$\rightarrow$$K$-operator (without dimensional loss), while the strategy proposed in this article, to provide a computational example of construction of brain spaces, is: ROIs$\rightarrow$ MDS $\rightarrow$positioning of $\mathcal{G}$s as points in the brain space, and subsequently the computation of the $K$-operator within the reduced space.
}

\clearpage
\newpage

\section{Conclusions}

In this article, we introduced the concept of brain spaces, where each point is the connectivity matrix of a 
connectome. The closer the points within the space, the more similar the connectomic configurations between 
them. Such a representation allows a quantitative analysis of similarity and dissimilarity between brains in 
states corresponding to specific diseases, or oscillating behaviours between healthy and diseased states, as 
in the case of mood disorders or epilepsy. We also quantitatively defined a path connecting two points of the 
space, relating it to the variation of connectivity between pairs of brain regions. In particular, the path 
from a healthy state to a diseased one, or between two diseased states as a disease progression, is the 
result of the action of the $K$-operator, recently proposed as a mathematical operator reproducing the key 
features of a disease, modifying the weights of the connections between brain regions.
\textcolor{black}{In this research, we chose to focus on functional connectivity, and thus the functional magnetic resonance imaging was adopted as the original source of our data. Future investigations will also include tractography, for a precise information on the anatomic alterations of the brain cortex. In fact, a more comprehensive investigation will also include a comparison between functional and anatomic alterations.}

Possible research developments concern an extension of the operator, including neurotransmitters, 
neurochemistry, and other biomarkers. This would allow a more comprehensive representation of the operator, 
as its action on different levels of the brain, and of other parameters influencing the brain states. 
Considering the healing process as the inverse of disease progression \cite{frontiers_paper}, an action on 
de-inflammation and neurochemistry can have a decisive role in the connectomic healing. 


To generalise the $K$-operator including other parameters, we could refer to the work presented in \cite{fard}, 
which shapes healthy and diseased states by using networks of multi-omic representations.

However, to advance this segment of research, we need a source of data on neurochemistry inflammation 
that can be translated into matrices.

Another issue to be addressed is stability, and more precisely, how to keep a brain's state within the subspace ``healthy'', allowing only fluctuations within a certain range.
\textcolor{black}{In particular, we have assumed, for what concerns the subspace of healthy brain states, a basin of attraction of brain states. The stability with respect to such a basin characterises a healthy brain, whose states are not pathological, or do not involve any pathological state. A process of complete healing from a neuropsychiatric disease involves the progressive disappearance of pathological states, and a permanence of the brain, after a certain time point, within the basin of attraction of healthy states. However, the precise definition of such a basin requires further experimental investigation.}


\section*{Declarations}

\subsection*{Ethical Approval} Not applicable.

\subsection*{Funding} This paper was developed within the project funded by Next Generation EU -- ``Age-It -- Ageing well in an 
ageing society'' project (PE0000015), National Recovery and Resilience Plan (NRRP) -- PE8 -- Mission 4, C2, 
Intervention 1.3, CUP B83C22004880006.

The research by P.F. was supported by the European Union within the REFRESH project -- Research Excellence 
For Region Sustainability and High-tech Industries ID No. CZ.10.03.01/00/22 003/0000048 of the European Just 
Transition Fund.

\subsection*{Availability of data and materials}
\textbf{Data availability.} Data concerning Alzheimer-Perusini's patients can be downloaded upon request from the Alzheimer's Disease
Neuroimaging Initiative (ADNI) database (\url{https://adni.loni.usc.edu}). According to the guidelines, we 
state that ``as such, the investigators
within the ADNI contributed to the design and implementation of ADNI and/or provided data but did not 
participate in the analysis or writing of this report. A complete listing of ADNI investigators can be found 
at:
\url{https://adni.loni.usc.edu/wp-content/uploads/how_to_apply/ADNI_Acknowledgement_List.pdf}.''

Data concerning Parkinson's disease patients can be downloaded upon request from Parkinson’s Progression 
Markers Initiative
(PPMI) dataset, with the following licenses and restrictions: ``Investigators seeking access to
PPMI data must sign the Data Use Agreement, submit an Online Application and comply with
the study Publications Policy. Requests to access these datasets should be directed to PPMI,
\url{https://ida.loni.usc.edu/collaboration/access/appLicense.jsp}.''

Data concerning the schizophrenic patient can be downloaded from the Center for Biomedical Research 
Excellence (COBRE) dataset, from the webpage \url{https://www.nitrc.org/pro jects/schizconnect/}.

\textbf{Code availability.} The code to compute the $K$-operator, given DICOM files as input for two different time points, can be 
accessed at \url{https://github.com/medusamedusa/K_operator_parkinson} (DOI 10.5281/zenodo.14162649), 
concerning in particular patient B, \#100006, atlas AAL3, and \url{https://github.com/medusamedusa/AAL3_K-operator_AD};
\textcolor{black}{ and the detailed discussion of the definition of the operator is proposed in \cite{PD_paper}.}

The code to compute the brain space can be accessed at \url{https://github.com/medusamedusa/BrainSpaces} and \url{https://codeberg.org/medusamedusa/BrainSpaces.git}, DOI: \url{https://doi.org/10.5281/zenodo.15302779}.


\bibliographystyle{plain}
\bibliography{sample}

\end{document}